# Surface properties of Co$_2$MnAl Heusler alloy

Amar Kumar[1], Sujeet Chaudhary[1*], and Sharat Chandra[2*]

[1]Thin Film Laboratory, Department of Physics, Indian Institute of Technology Delhi, New Delhi 110016, India

[2]Material Science Group, Indira Gandhi Centre for Atomic Research, Homi Bhabha National Institute, Kalpakkam, Tamil Nadu-603102, India

*Corresponding Authors: sujeetc@physics.iitd.ac.in and sharat@igcar.gov.in

## Abstract

Using the plane-wave pseudopotential method within the framework of density functional theory, Co$_2$MnAl (100), (110), and (111) surfaces with different atomic terminations have been studied in the context of some key spintronics properties, *viz.*, surface energy, half-metallicity, magnetization, and magnetic anisotropy. The present study reveals that the MnAl-(100), Co-Al-(111), and Al-(111) surfaces exhibit negative surface energies over a wide range of chemical potentials, indicating their strong structural stability. The MnAl-(100), CoCoMnAl-(110), and Co-Mn-(111) surfaces maintain the nearly half-metallic nature like the bulk-Co$_2$MnAl, while this nearly half-metallic nature even improved for the Al-(111) surface. In contrast, the rest of the considered surfaces – CoCo-(100), Co-Al-(111) and Mn-(111) surfaces – display the strong metallic nature. Magnetization is enhanced for most surface configurations, except for Al-(111), where it decreases due to reduced moments of the exterior atoms. Regarding magnetic anisotropy, only the MnAl-(100) and Co-Mn-(111) surfaces exhibit the positive magneto-crystalline anisotropy of ~0.23 and ~0.33 mJ/m$^2$, respectively. All these findings suggest that the Co-Mn-(111) and MnAl-(100) surfaces are quite appealing for spintronics applications, considering the structural stability, electronic properties, and magnetic anisotropy.

# 1. Introduction

Spintronics is an emerging research field aiming to provide next-generation nanoelectronics devices through the implementation of various magnetic configurations, such as - spin valves, magnetic tunnel junctions (MTJs), spin transistors, *etc*. Although each spintronics device requires specific optimized features for its constituting layers, almost all of them require at least one layer of ferromagnetic (FM) materials with very high spin polarization ($P$) and very high curie temperature ($T_c$). Amidst the different FM materials with such properties, Heusler alloys (HAs), particularly those based on the 3$d$-transition metals-based stand out as one of the most popular choice. Not only do they fulfill the criteria for high $P$ (≈100%, the perfectly or nearly half-metallic ordered structure) and high $T_c$; but they also display the preferable results for a large set of other physical properties, which are essential for most of the spintronic devices, such as - low damping parameter, moderate magnetization and magnetic anisotropy, *etc*. These attributes establish HAs as an adequate choice for FM in spintronics devices [1,2].

However, achieving such beneficial outcomes for HAs in experiments is challenging due to the presence of structural imperfections. Here, the structural imperfections come into the picture from the experimental limitations, primarily from the deposition methodologies and the application regimes of HAs. Deposition methodologies often lead to the growth of various point defects, sometimes accompanied by lattice deformations for the FM material. On the other hand, when HAs are utilized in the thin film regime, the surface termination impact becomes significant, alongside the point defects and lattice deformations. Notably, the thin films, which are the basis for the device fabrication, essentially lead to different surface terminations for the material utilized in device fabrications. Since bond reformations occur at the surface terminations, material properties deviate significantly from those of bulk geometries. Therefore, whenever HAs are used in the thin film regimes, studying the surface properties of HAs along with study for the impact of point defects is crucial for optimizing the device design [3–8].

Among the extensive family of Heusler alloys, $Co_2MnAl$ is a key material for spintronics applications. It exhibits a very high $P$ of ≈ 75 % (theoretical value for L2$_1$ ordered structure, nearly half-metallic nature), ≈ 65 % (experimental value) [9–11], and a high Curie temperature of ≈ 720 K - 750 K [7]. Remarkably, these properties are comparable to those of CoFeB (which is the most widely utilized material for spintronics to date), for which a maximum polarization of ≈ 65% [12,13 and refs. therein] and a Curie temperature of 750 to 1000 K [14,15] have been reported so far, to the best of our knowledge. Such high spin polarization and Curie temperature help to establish $Co_2MnAl$ as a very promising candidate for spintronics applications. Regarding the study of the structural imperfections in $Co_2MnAl$, there are many studies for the point defects and lattice deformations [7,8,16–18]. However, despite its exceptional electronic and magnetic properties and the significance of studying the surface characteristics of a candidate material, the surface properties have not been much explored for $Co_2MnAl$ compared to the other Co- and Mn-based HAs, and only a few reports are available for the electronic and magnetic properties of $Co_2MnAl$ surfaces. These few studies focus on exploring the electronic and magnetic properties of $Co_2MnAl$ (100) surfaces (namely, spin polarization, magnetization, and the exchange constants), and show that the MnAl-terminated and Co-terminated $Co_2MnAl$ (100) surfaces exhibit nearly half-metallic and strong metallic behavior, respectively. On the other hand, for both kinds of terminated (100) surfaces, significantly increased magnetization is observed as compared to the bulk geometry [19,20].

Therefore, the literature clearly highlights the need for a comprehensive study of the different electronic and magnetic properties of $Co_2MnAl$ surfaces, particularly $Co_2MnAl$ surfaces with various surface orientations and atomic terminations. The state-of-the-art first principles-based methods can be utilized to evaluate the impact of surface formation on various desired physical properties prior to the actual design of materials via expensive experiments. Moreover, the first-principles results can also be used to control or manipulate these properties according to specific requirements, making them a very useful tool for material design. Therefore, in the present study, we aim to investigate the impact of surface terminations on the various structural, electronic, and magnetic properties, using the first principles-based density functional theory (DFT) calculations. For surface modeling, we have considered only three different high-symmetry surfaces in the present study: - (100), (110), and (111) - along with all possible atomic terminations for each surface orientation. While a more rigorous study combining the surface terminations with point defects or lattice deformations, could be closer to realistic situations and hence the more desirable, the available computational constraints limited the present study to epitaxial surface geometries. Furthermore, to facilitate the experimental utilization of these results for heterostructure development later, the thicknesses of the slabs for modeling the $Co_2MnAl$ surfaces are chosen within experimentally appropriate regimes, as discussed later in the Results and Discussion section (vide the beginning of Section 3). Further details about the surface modeling are provided in Section-2 (Computational Details) and Section-3 (Results and Discussions). The structural properties for surface slabs are discussed in Subsection 3.1, and Subsection 3.2 encloses the result for the impact of surface terminations on the electronic and magnetic properties of $Co_2MnAl$. The consequence of surface terminations- the magneto-crystalline anisotropy, is discussed in Subsection 3.3. Finally, all results are summarized in the concluding section- Section 4.

## 2. Computational details

First principle-based density functional theory (DFT) calculations are performed using QUANTUM ESPRESSO code to gain insight into the physical properties of different $Co_2MnAl$ surfaces [21,22]. For the atomic potentials, the pseudopotential from the publicly available repository (PSlibrary) with valence electronic configuration of Co ($3s^2 4s^2 3p^6 3d^7$), Mn ($3s^2 4s^2 3p^6 3d^5$), and Al ($3s^2 3p^1$) are used [23]. For treating electronic exchange and correlation (XC) interactions, Perdew-Burke-Ernzerhof functional within the generalized gradient approximation (GGA) XC functional is employed [24]. For the surface relaxations and electronic bands calculations of all surface slabs, the Brillouin zone is sampled according to the Monkhorst-Pack scheme, with a *k*-point mesh equivalent to 7×7×7 *k*-points for the L2$_1$-ordered structure of Co$_2$MnAl. The maximum energy of the plane wave basis set is limited to 150 Ry. For achieving the minimum energy (optimized) configurations for the surface slabs, the five atomic layers closest to the vacuum in each slab are allowed to relax in the $\hat{z}$ (vacuum) direction using the Davidson iterative diagonalization method, with an atomic force convergence criterion of 10$^{-3}$ Ry/bohr. For the force convergence, all components of all forces on all atoms must be below the convergence threshold. Self-consistency is achieved when the total energy difference between two consecutive steps is less than 10$^{-6}$ Ry. All calculations are performed within the scalar relativistic framework for computational efficiency, considering the fact that the heavy elements are not present in the studied structures. In other words, the spin–orbit coupling (SOC) is not included in the calculations, and only collinear electronic spins are considered; as SOC effect on the electronic and magnetic properties of Co$_2$MnAl has already been established to be negligible [25]. Furthermore, while it is often observed that Hubbard '*U*' correction to the GGA functionals is needed for a more accurate description of *d*-bands, we have not utilized GGA+*U* method for the electronic structure calculations of Co$_2$MnAl. This is due to the fact the 'DFT+*U*' method did not give better predications of magnetic moments (total as well as atomic) for Co$_2$MnAl, as observed in our previous study and in many other references as well [8,26–28]. In Section-4, the magneto-crystalline anisotropy is calculated using the magnetic force theorem as implemented in QUANTUM ESPRESSO code [29], and fully relativistic pseudopotentials are utilized for these calculations. Further details regarding the adopted method of magnetic anisotropy calculation and computational parameters are given in the relevant section.

## 3. Results and Discussions

Before discussing the simulation results for the various physical properties of different Co$_2$MnAl surfaces, let us briefly address the slab modeling of Co$_2$MnAl surfaces. In the literature, (100) surfaces are often the preferred choice for investigating the impact of surface terminations on various physical properties of HAs. This preference is likely due to the high symmetry of the HAs-(100) plane compared to the other atomic planes, as well as its excellent lattice matching with a wide variety of mainstream materials – such as MgO, MgAl$_2$O$_4$, GaAs, Si, SiO$_2$ – which facilitates the easy modeling of spintronic heterostructures. In contrast, other high symmetry surfaces, such as (110) and (111), have received comparatively less attention, as reflected in many studies describing the surface and interfacial properties of 3*d*-based HAs [6,30–35]. Some other potential reasons for this preference might include concerns about the structural instabilities and the quenching of material properties in other surface planes. However, all these assumptions need to be verified for the specific material, particularly those with uncommon band structures, such as Co$_2$MnAl. Moreover, the high symmetries and lattice parameters of the (110) and (111) surfaces of cubic 3*d*-transition-metal-based HAs suggest that they can also offer various unique surface terminations and numerous possibilities for heterostructure formation with other materials. As such, these surfaces could result in some exclusive physical properties due to their distinct atomic terminations and orientations, differing from those of the (100) surfaces. This speculation about the (110) and (111) surfaces is also supported by some recent theoretical studies, which show that (110) and (111) surfaces of different HAs are also significantly important due to their unique physical properties and suitability for forming heterostructures with various materials [36–40]. This indicates that studying the physical properties of different high-symmetry surfaces with various orientations and atomic terminations would be highly beneficial for advancing the literature. Therefore, in the present work, we have attempted to explore the impact of surface terminations, with different planar orientations – (100), (110), and (111) – along with distinct atomic terminations, on the electronic and magnetic properties of $Co_2MnAl$. Additionally, the structural properties of these surfaces have been studied through surface relaxations and surface energies and the induced magnetic anisotropy resulting from the surface terminations is also analysed. Since the previous studies on $Co_2MnAl$ surfaces have been limited to analyzing the magnetization and half-metallicity of $Co_2MnAl$-(100) surfaces, hence, this work holds potential significance to guide the material design for spintronics devices using different $Co_2MnAl$ surfaces. The modeling methodology and naming conventions for the studied surfaces are as follows:

All surface slabs are modeled using the ideal ordered structure of $Co_2MnAl$ ($L2_1$-ordered structure with 5.69 Å lattice parameter). The symmetric slabs, with an odd number of atomic-layers and equivalent atomic terminations at both ends, are used to represent the $Co_2MnAl$ surfaces to avoid the surface dipole creations. Additionally, a 10 Å vacuum layer on each side of slab is imposed to prevent the unphysical overlap of the atomic wave functions between the periodic images of surface slabs. The ideal epitaxies of the surface slabs are utilized for studying the physical properties of surfaces. As in the [100] direction; the $L2_1$-ordered $Co_2MnAl$ structure comprises the alternating atomic planes consisting of the CoCo-atoms and MnAl-atoms, two possible kind of atomic terminations can exist for $Co_2MnAl$ (100)-oriented surfaces – the first one is CoCo-terminated (Figs. 1(I) and 1(a)), and the second one is MnAl- terminated (Figs. 1(II) and 1(b)). For convenience, the CoCo-terminated $Co_2MnAl$ (100) surface is abbreviated as CoCo-(100) surface, while the MnAl-terminated $Co_2MnAl$ (100) surface is abbreviated as MnAl-(100) surfaces. Whereas in the [110] direction, there exists only one type of atomic plane with the same stoichiometry as the native compound (*i.e.*, $Co_2MnAl$), leading to a single kind of (110)-surface which is termed as CoCoMnAl-(110) surface (Figs. 1(III) and 1(c)). On the other hand, in the [111] direction, there are alternating mono-atomic planes of Co, Mn, and Al atoms with a Co–Mn–Co–Al stacking sequence, resulting in four different types for $Co_2MnAl$ (111) surfaces, when including the subsurface layers also. With the atomic stacking from top to bottom direction (or in the inward direction ($-\hat{z}$) from the slab topmost layer), are referred as:

- Co-Mn-Co-Al stacking sequence, termed as Co-Mn-(111) surface (Figs. 1(IV) and 1(d))
- Al-Co-Mn-Co stacking sequence, termed as Al-(111) surface (Figs. 1(V) and 1(e))
- Co-Al-Co-Mn stacking sequence, termed as Co-Al-(111) surface (Figs. 1(VI) and 1(f))
- Mn-Co-Al-Co stacking sequence, termed as Mn-(111) surface (Figs. 1(VII) and 1(g))

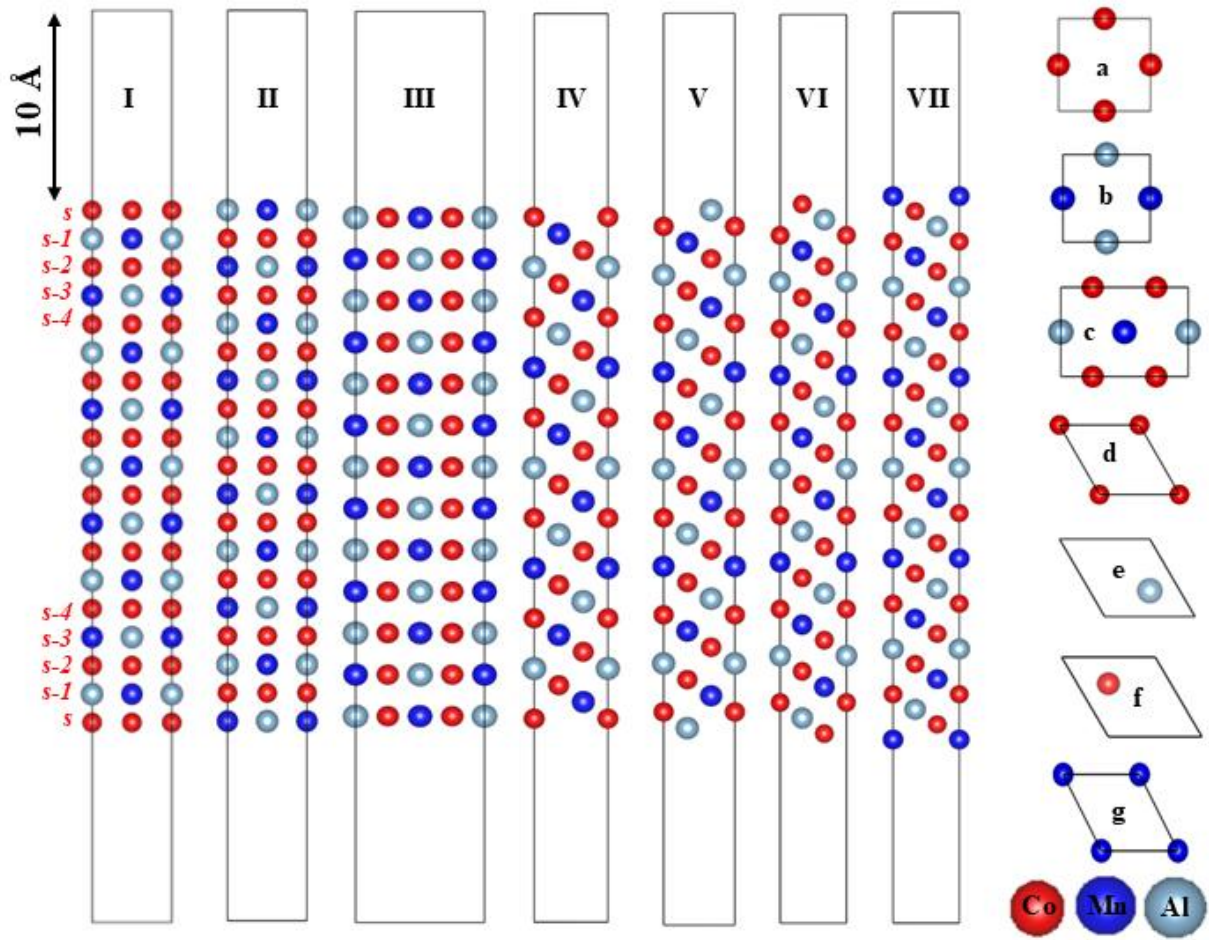

**Figure 1:** Side- and top-views of the different $Co_2MnAl$ surface slabs for: (I) and (a) CoCo-(100) surface, (II) and (b) MnAl-(100) surface, (III) and (c) CoCoMnAl-(110) surface, (IV) and (d) Co-Mn-(111) surface, (V) and (e) Al-(111) surface, (VI) and (f) Co-Al-(111) surfaces, (VII) and (g) Mn-(111) surface. Atomic layers towards slab direction, from the top of surface, are labelled as s, (s-1), (s-3), etc. See text for more details on the slab modeling.

Here, for the Co-Mn-(111) and Co-Al-(111) surfaces; the 1$^{st}$ atoms belong to the surface layer (s), whereas the 2$^{nd}$ atom belongs to first sub-surface layer (s-1), as shown in Fig. 1. Similarly, for other slabs also, the s and (s-1) layer represent the surface and sub-surface layer, respectively. The other layers - (s-2), (s-3), …, (s-n) – correspond to the 2$^{nd}$, 3$^{rd}$, …, n$^{th}$ beneath layer from the top of surface slabs, in the inward direction.

Regarding the choice of slab thickness (and, consequently, the number of atomic layers), the surface slabs are generally chosen to be thick enough to replicate bulk-like properties in the central layers, such as atomic magnetic moments (AMMs) and partial density of states (PDOS), unless specified otherwise for a particular application. Additionally, the outermost layers (those near slab terminations) should exhibit similar physical and chemical characteristics, including surface reconstruction and bond lengths, even if the slab thickness changes. In general, thicker slabs yield more accurate results for surface terminations but also entail higher computational costs. However, once a slab exceeds a certain critical thickness (which depends on the material and surface orientation), it becomes sufficient for simulating electronic properties. Therefore, an optimal slab thickness should be chosen carefully to balance numerical accuracy and computational efficiency. Notably, in most experiments related to spintronics devices reported in literature, the FM layer thickness lies within 1-3 nm. On the other hand, for simulations, slabs with ~2 nm thickness are commonly chosen for studying the physical properties of various surfaces. Considering that, 19 monolayer (ML) slabs for both kinds of (100) surfaces with a resulting thickness of ~2.5 nm, 13 ML slab for (110) surface with a resulting thickness of ~2.4 nm thickness, 31 ML slab for Mn-Co-(111) surface with ~2.46 nm thickness, 33 ML slab for Al-(111) surface with ~2.62 nm thickness, 35 ML slab for Al-Co-(111) surface with ~2.79 nm thickness, 37 ML slab for Mn- (111) surface with ~2.95 nm thickness are considered. Based on the atomic DOS and AMMs for the middle- and interior-layers' atoms, and bond lengths of the near surface atoms of all the considered slabs, it is confirmed that the chosen slab thicknesses are sufficient to simulate the surface physical properties accurately. It is significant to point out here that for a family of planes, simulations are typically performed with the same thickness or with the same number of atomic layers, to enable comparison of results. However, to maintain identical morphologies at both slab ends of different (111) $Co_2MnAl$ surfaces, varying numbers of layers naturally arise depending on their atomic terminations due to the bulk crystal structure of $Co_2MnAl$. Since the thickness in each case is sufficient to accurately mitigate the impact of surface terminations, the results can be reliably compared for the near surface layers. The different structural parameters - in-plane lattice parameters, number of atoms per layer, interlayer distances, and resulting space group (for (1×1) slab unit-cell) for each slab are summarized in Table 1. Furthermore, as a noteworthy informational point, the atoms inside the layer interchange (swap) their positions (Mn and Al) with repetitions at the (S±2), (S±1), and (S±4) layers for the (100), (110), and (111) surfaces, respectively (cf. Figure 1). The selected in-plane lattice vectors align with the crystal's high symmetry directions and preserve the inherent symmetry of the bulk crystal for the innermost layers of the surface slabs. The following sub-sections describe the various physical aspects of the above-mentioned $Co_2MnAl$ surfaces:

**Table 1:** Structural details of the $Co_2MnAl$ surface slabs. The optimized lattice parameter of $Co_2MnAl$ (i.e., $a = 5.69$ Å) is utilized for modeling the all-surface slabs.

| Surface | Atoms per Layer | Interlayer Distance | LatticeVectors (U, V, W) | In-Plane Lattice Parameters | Space group |
|---|---|---|---|---|---|
| (100) | 2 | $a/4$ | (010), (001), (100) | ($a/\sqrt{2}$, $a/\sqrt{2}$) | P4/NMM (#129) |
| (110) | 4 | $a/2\sqrt{2}$ | (001), (1$\bar{1}$0), (110) | ($a/\sqrt{2}$, $a$) | PMMM (#47) |
| (111) | 1 | $\sqrt{3}a/10$ | (1$\bar{1}$0), (01$\bar{1}$), (111) | ($a/\sqrt{2}$, $a/\sqrt{2}$) | P$\bar{3}$M1 (#164) |

### 3.1 Structural properties of $Co_2MnAl$ surfaces

The structural properties of different $Co_2MnAl$ surfaces have been studied through surface relaxations and surface free energies. Since the breaking of crystal symmetry, a distinct atomic structure appears for the near-surface atomic layers of the surface slabs, differing from the bulk geometry of the native compound. Thus, surface relaxations may occur, and the surface could face considerable reconstruction, as indicated by the previous studies [41,42]. Thereby, before calculating the physical properties of the surface slabs, it is crucial to capture the impact of surface terminations on the slab geometry, most likely *via* the atomic position relaxations; especially for the atomic layers in the proximity of slab ends where the surface formation impact will be most dominant. Consequently, the atoms of the five atomic layers closest to the slab terminations (from the s to the (s-4) layer) of all studied surface slabs are allowed to relax. For the remaining atomic layers, the atoms are kept fixed, as surface formation is expected to have a minimal impact on them, leading to negligible relaxation. The surface relaxation results are shown in Figure 2, showing the changes in the atomic positions in the orthogonal direction to the surface planes (along the vacuum direction) after relaxation. The relaxations results are also tabulated in Table S1 of the supplemental material for numerical values, showing the changes in the atomic positions ($\Delta z$) in the orthogonal direction to the surface planes (along the vacuum direction) after relaxations (column 3), and their percent value relative to the interlayer distances (column 4). In Table S1, the $\Delta z$ is calculated as ($z_f$ - $z_i$) with $z_f$ and $z_i$ are the slab atoms' z-coordinate after and before relaxation, respectively. Since the relaxation amplitude for the (s-4) layer is small, the corresponding data for (s-4) layer for slabs is not included in Table S1.

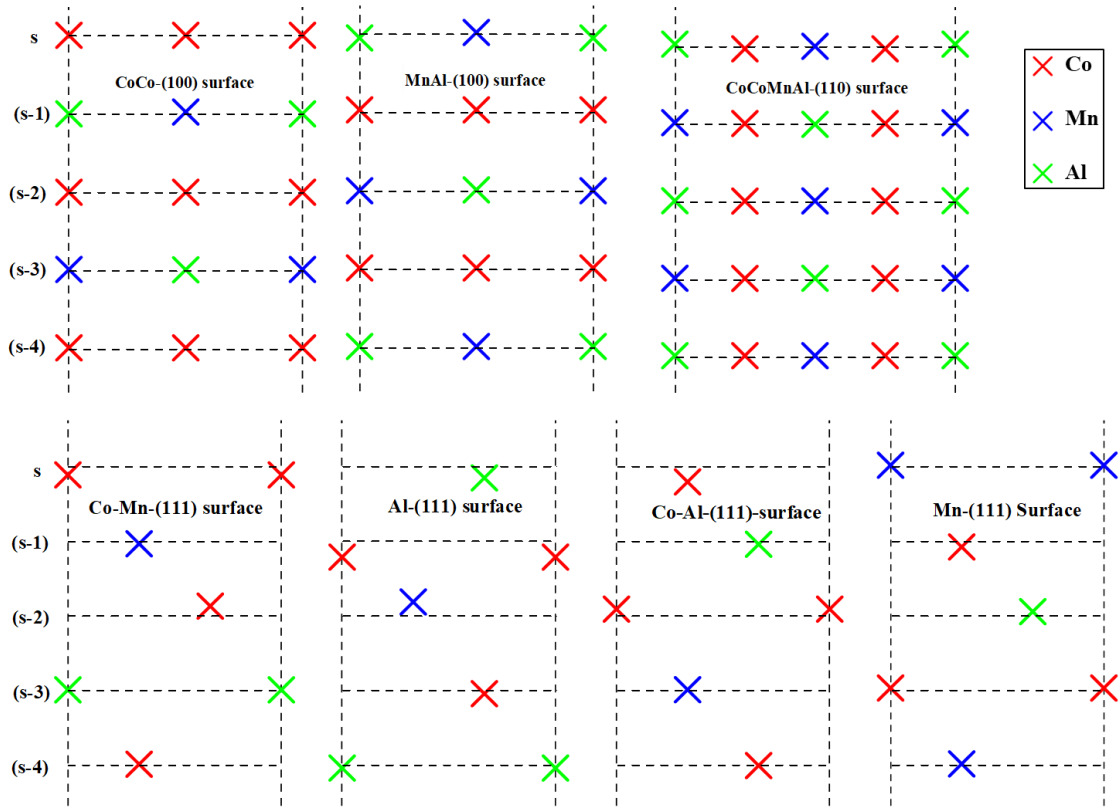

**Figure 2**: Schematic representation of the relaxed atomic positions of different $Co_2MnAl$ surface slabs. The vertical axis shows the *z* component of atomic positions. The dashed lines indicate the atomic planes within the non-relax slabs for the surface and other subsurface layers. The red, blue and green cross symbols (×) correspond to the relaxed-atomic positions of Co, Mn and Al in the slab.

After relaxation, it is observed that in the near-surface atomic layers of all slabs, atoms within a particular layer can move in opposite directions when multiple atomic species are present. Some atoms exhibit inward movement toward the slab (indicated by negative $\Delta z$), while other atoms shift toward the vacuum (indicated by positive $\Delta z$), resulting in non-planar atomic layers. Also, the impact of surface relaxations is significant in the near surface layers and decreases progressively with depth into the slabs, as indicated by the $|\Delta z|$ values. As observed from Figure 2 and Table S1, atomic relaxations are minimal for the (100) and (110) surfaces, compared to the (111) surfaces. Specifically, for both (100) surfaces, atomic position changes occur up to the second subsurface (s-2) layer, whereas for the (110) surface, changes are confined to the first subsurface (s-1) layer. Moreover, for these surface slabs, the changes in atomic positions are less than 5% relative to the interlayer distances. In contrast, all (111) surfaces experience significant relaxation due to their lower symmetry (space group #164). Their smaller interlayer spacing, compared to the other surface slabs, facilitates greater atomic wavefunction overlaps along the vacuum axis, resulting in larger relaxations (apparent $|\Delta z|$ values) that extend beyond the (s-2) layers. For these surfaces, the relaxation amplitude gradually diminishes deeper into the slab and atomic relaxations are observed up to the (s-4) layer. Therefore, it can be concluded that the minimum surface reconstruction occurs for the (100) and (110) surfaces, while the (111) surfaces undergo notable relaxations.

Microscopically, changes in the atomic positions of near-surface layers, and consequently surface reconstructions, are primarily driven by redistribution of atomic charges among near-surface atoms, along with the altered ionic-potential energy landscape near the slab geometry's ends. The modifications in the ionic potential energy landscape near slab terminations, and atomic charges redistribution in the near-surface region, result from the combined effects of surface terminations and differences in electronegativities of the constituent atoms. In turn, the altered ionic potential leads to changes in the total energies of near-surface atoms, while residual-electrons from broken bonds at surface terminations redistribute among near-surface atoms based on their electronegativities and atomic masses. In combination, these factors result in different interatomic forces between near-surface atoms, leading to new equilibrium positions.

Additionally, for the (100) and (110) slabs, the layers contain multiple types of atoms. Therefore, the atom-specific displacement in the opposite direction, driven by differing interatomic forces, leads to uneven layers for these slabs. Notably, in these slabs, beyond the third sub-surface layer, the relaxation amplitude becomes negligible, and the resulting non-planarity of the atomic layers is minimal, making the deeper layers appear planar. In contrast, for the (111) surfaces, where each layer contains a single type of atom, the interlayer slab separation becomes uneven due to the unequal displacement of constituent atom. To sum up, it can be concluded that for the (100) and (110) surfaces, minor relaxations occur, and the near surface atomic layers (particularly, surface and 1$^{st}$ subsurface layers) become slightly uneven after relaxation. Conversely, the (111) slabs experience the substantial relaxation extending to multiple layers from surface, leading to uneven interlayer separation up to the 3$^{rd}$ subsurface layer. Additionally, for a given orientation of the surface slabs, due to the nearly identical degree of relaxation in the near-surface layers, it is difficult to distinguish between the atomic-termination with the highest and lowest reconstruction for a particular orientation. Although, a better quantitative representation of surface relaxation could involve analyzing the changes in bond lengths among the near-surface layer atoms, this aspect is not addressed in the present study.

Furthermore, to evaluate the relative stability of different surfaces, the surface free energy (which is also known as the surface formation energies) is also calculated for all surfaces, and shown in Figure 3. For calculating the surface free energies, the following equation is utilized –

$$\gamma = \frac{1}{2A}\left[G - \sum_i (N_i \mu_i)\right] \quad (1)$$

Here, G is the Gibbs free energy of a particular surface slab, N represent number of i$^{th}$ atoms, $\mu_i$ is chemical potential of the i$^{th}$ atom in the corresponding slab, and A is the in-plane area of the same surface slab. For the calculation of surface energies using Eqn (1), the DFT-calculated total energy can be safely used for Gibbs free energies of surface slabs, like in other references, since the dynamic contributions to Gibbs free energy at sufficiently low temperatures is negligible [38,43–45]. Since the surface slabs have bulk-like central layers, and their surfaces are in the thermodynamic equilibrium with central bulk layers, the sum of chemical potentials would be same as of total bulk energy of Co$_2$MnAl. Thereby, the following equilibrium condition is imposed on the atomic chemical potential by default:

$$2\mu_{Co} + \mu_{Mn} + \mu_{Al} = G_{Co_2MnAl} \quad (2)$$

By utilizing the Eqns (1)-(2), the surface energies of the surface slabs can be calculated as the function of only two independent chemical potentials. Here, the chemical potential of Co and Mn are selected for the calculation of surface energies, because the evaporation of Co and Mn is simpler than that of Al in experiments, and therefore slightly nonstoichiometric deposition of Co and Mn in the grown film can be easily achieved. Here, the chemical potentials of Co and Mn are the variables dependent on the thermodynamic conditions. The boundaries of these potentials are defined as:

$$\frac{1}{2}(G_{Co_2MnAl} - G_{MnAl}) \leq \mu_{Co} \leq G_{Co} \tag{3}$$

$$G_{Co_2MnAl} - (G_{CoAl} + G_{Co}) \leq \mu_{(Mn)} \leq G_{(Mn)} \tag{4}$$

Here, $G_X$ represents the ground state DFT energies of X-compound. Eqn (3) (Eqn (4)) indicates that if Co (Mn) concentration is lowered, the sample would decompose into bulk MnAl (bulk CoAl + bulk Co) and elementary Co (Mn) droplets. Whereas, under the elemental rich conditions, bulk- Co (Mn) will be deposited. For the calculation of ground state energies of compounds/elements in Eqns. (3) and (4), the ferromagnetic (FM) hexagonal close-packed Co, [001] antiferromagnetic-face-centered cubic (*fcc*) Mn, and nonmagnetic (NM)–*fcc* Al, FM L2$_1$-ordered Co$_2$MnAl, NM-cubic CoAl, and tetragonal-FM MnAl phases are utilized. For all these compounds, the total energies are calculated with similar computational approaches as for the surface slabs. Notably, under Mn-poor conditions, the decomposition of Co$_2$MnAl into three components—bulk CoAl, bulk Co, and elemental Mn droplets—results in a wider range of atomic chemical potential for Mn compared to Co. The calculated surface energies are displayed in Figure 3. From Figure 3, it is clear that out of seven surface configurations, only three surfaces (MnAl-(100), Co-Al-(111), and Al-(111) surfaces in Figure 3(a)) show the negative surface energies over the entire range of chemical potential, indicating their highest stabilities. On the other hand, the CoCo-(100) and CoCoMnAl-(110) surfaces show small positive surface energies, suggesting their low stability for these surfaces. The remaining surface configurations exhibit very highly positive surface energies, reflecting their lowest stability.

By calculating and comparing the surface energies of the slabs, along with their relative stabilities, their feasibility of formation can also be assessed. The surfaces with negative surface energies—MnAl-(100), Al-(111), and Co-Al-(111)—demonstrate the potential for spontaneous growth in experiments. In contrast, slabs with lower positive surface energies (CoCo-(100) and CoCoMnAl-(110)) are inherently more stable and thus more likely to be experimentally fabricated. Conversely, surfaces with higher positive surface energies- the Mn-(111) and Co-Mn-(111) surfaces, are less stable and less likely to be realized experimentally. The magnitudes of calculated surface energies also align with their structural relaxation amplitudes, as discussed in the preceding paragraph. The (110) surface shows the smallest relaxation, corresponding to smaller surface energies. In contrast, the (111) surfaces exhibit the largest relaxation, resulting in the largest surface energy amplitude. To sum up, depending on the matching of in-plane lattice parameters of the surface slabs with adjacent layers, the MnAl-(100) surface within the {100}-planes, the CoCoMnAl-(110) surface within the {110}- planes (the only one with <110> orientation), and the Al-(111) and Co-Al-(111) surfaces within the {111}-planes are the most stable and most preferable for experimental growth. While the other surfaces within each orientation group are less likely to form compared to these surface terminations, they can still be experimentally achieved through non-equilibrium thin-film growth methods, such as sputtering or molecular beam epitaxy. Finally, it should be noted that the Gibbs free energies used to calculate the surface energies and the allowed range of chemical potential, are determined under absolute pressure and temperature conditions. Therefore, these results are valid only up to a finite temperature, beyond which dynamic contributions to the Gibbs free energies become non-negligible. For a similar material $Co_2MnSi$, this temperature is determined to be 400–500 K [45].

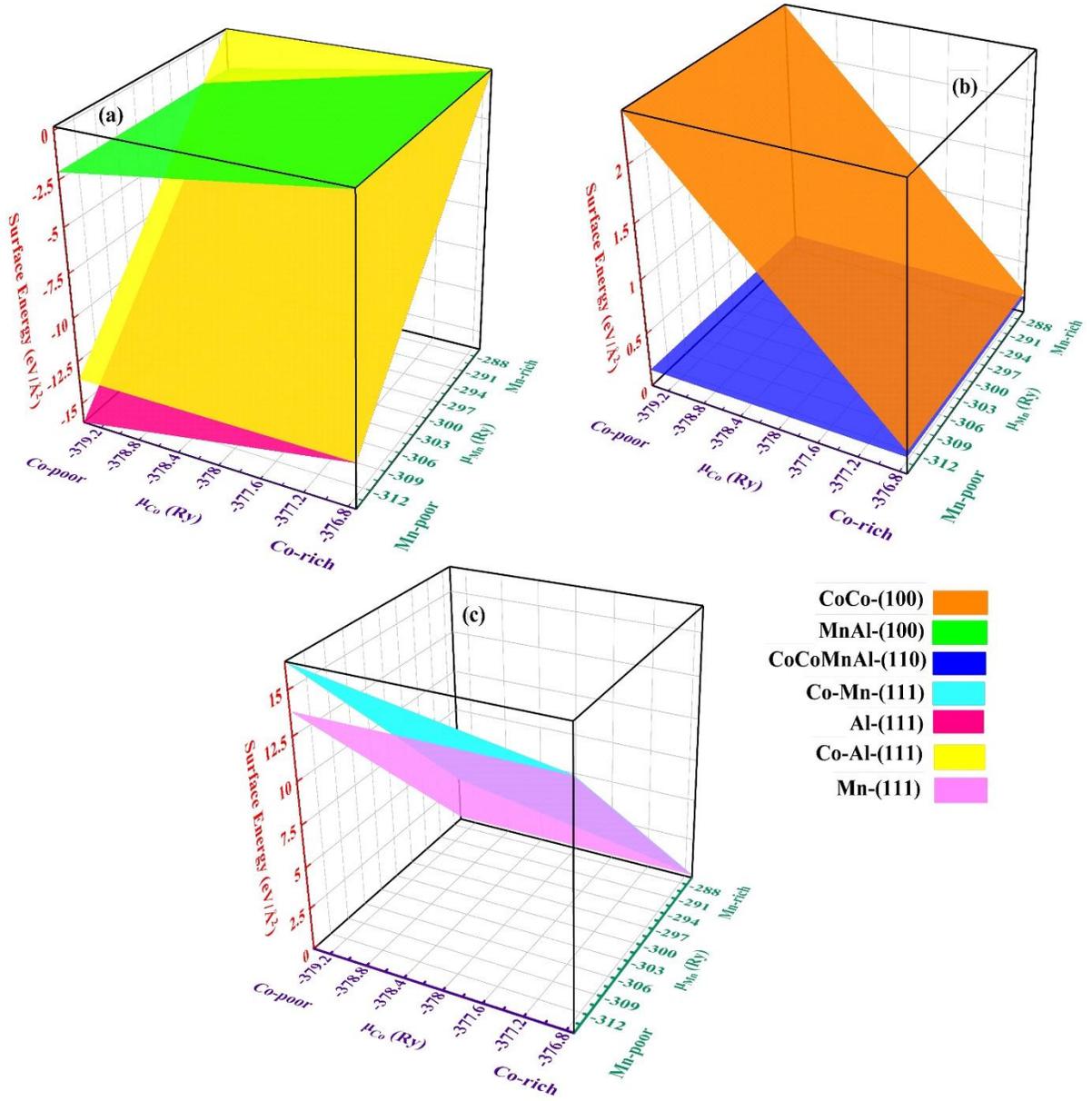

**Figure 3**: The calculated surface energy (eV/Å$^2$) of different surfaces as a function of chemical potential of Co and Mn – (A) surfaces with the negative surface energies - Co-Al-(111), Al-(111) and MnAl-(100) surfaces, (B) surfaces with small positive surface energies - CoCo-(100) and CoCoMnAl-(110) surfaces, and (C) surfaces with very high positive surface energies - Co-Mn-(111) and Mn-(111) surfaces. The surfaces are grouped into different figures according to their range of surface energies range for the plotting convenience.

### 3.2 Electronic and magnetic properties of $Co_2MnAl$ surfaces

Next, we discuss the electronic and magnetic properties of all considered surface slabs. To characterize the electronic properties of these surface slabs, the PDOS for surface slab atoms are calculated, as shown in Figure S1 of the supplemental material. To highlight the differences between the electronic nature of slabs and bulk $Co_2MnAl$, the atomic DOS are plotted along with the corresponding atomic DOS in the bulk $Co_2MnAl$. As illustrated in Figure S1, the PDOS for the near surface atoms (atoms of the near-surface layers) appear to be different from the corresponding PDOS of the bulk $Co_2MnAl$. Particularly, these changes in atomic DOS were observed up to second subsurface (s-2) layer, first subsurface (s-1) layer, and third subsurface (s-3) layer for the {100}, {110}, and {111} surface slabs, respectively. The changes are most pronounced for the slab top-layers' atoms (surface atoms) and gradually diminish as we move deeper into the slab. However, beyond the (s-3) subsurface layer, the changes in the PDOS for all slabs become minor, and the atoms almost replicate the PDOS similar as in bulk-$Co_2MnAl$. All these changes stem from the atomic relaxations of the near surface atoms, which, in turn, alter the electronic exchange interactions for these atoms and, consequently, result in the modified PDOS, as shown in Figure S1(a)-(g). Owing to these changes in PDOS of the near surface layers' atoms, the electronic nature of $Co_2MnAl$ in slab form (or $Co_2MnAl$ surface) is evidently differs from the near half-metallic nature of the bulk-$Co_2MnAl$. To summarize these PDOS changes, and to better identify the overall electronic nature of the surface slabs, the spin-resolved DOS projected around the surface region are plotted in Figure 4(a) for the top five atomic layers near the slab termination (from s to (s-4) layer). As these atomic layers capture almost all the surface formation influences on PDOS, the DOS plot in Figure 4(a) can also be referred to as the surface DOS plot for the corresponding slabs.

From Figure 4(a), it is clearly observed that the Al-(111) surface exhibits a nearly half-metallic nature, identified by the Fermi level ($E_F$) lying in pseudogap for minority spins along with finite DOS for majority spins at the Fermi level ($E_F$), like in the bulk native compound $Co_2MnAl$. More specifically, the Al-(111) surface exhibits an enhanced near-half-metallic nature compared to the bulk $Co_2MnAl$, due to greater majority spin DOS and wider pseudogap compared to bulk $Co_2MnAl$. On the other hand, with the formation of the CoCo-(100), Co-Al-(111) and Mn-(111) surfaces, the near half-metallic nature of bulk $Co_2MnAl$ is disrupted due to presence of significant DOS in pseudogap, and thereby aforementioned surfaces exhibit the strong metallic nature. Meanwhile, the remaining surfaces, i.e., MnAl-(100), CoCoMnAl-(110), and Co-Mn-(111), show the nearly half-metallic nature, similar to bulk-$Co_2MnAl$, but with few changes in the minority pseudogap. Due to the emergence of the minority states in the vicinity of $E_F$ (in the pseudogap) for these surface configurations, their near half-metallic nature is unstable, and the metallic behavior can be easily observed for them upon the slight variation in the structural or thermodynamic parameters. Apart from these changes, the DOS shape for the near surface atoms is also altered, with observation of sharper peaks in PDOS. These significant differences in DOS or PDOS imply that the surfaces will show subtle change in their transport properties from bulk-$Co_2MnAl$.

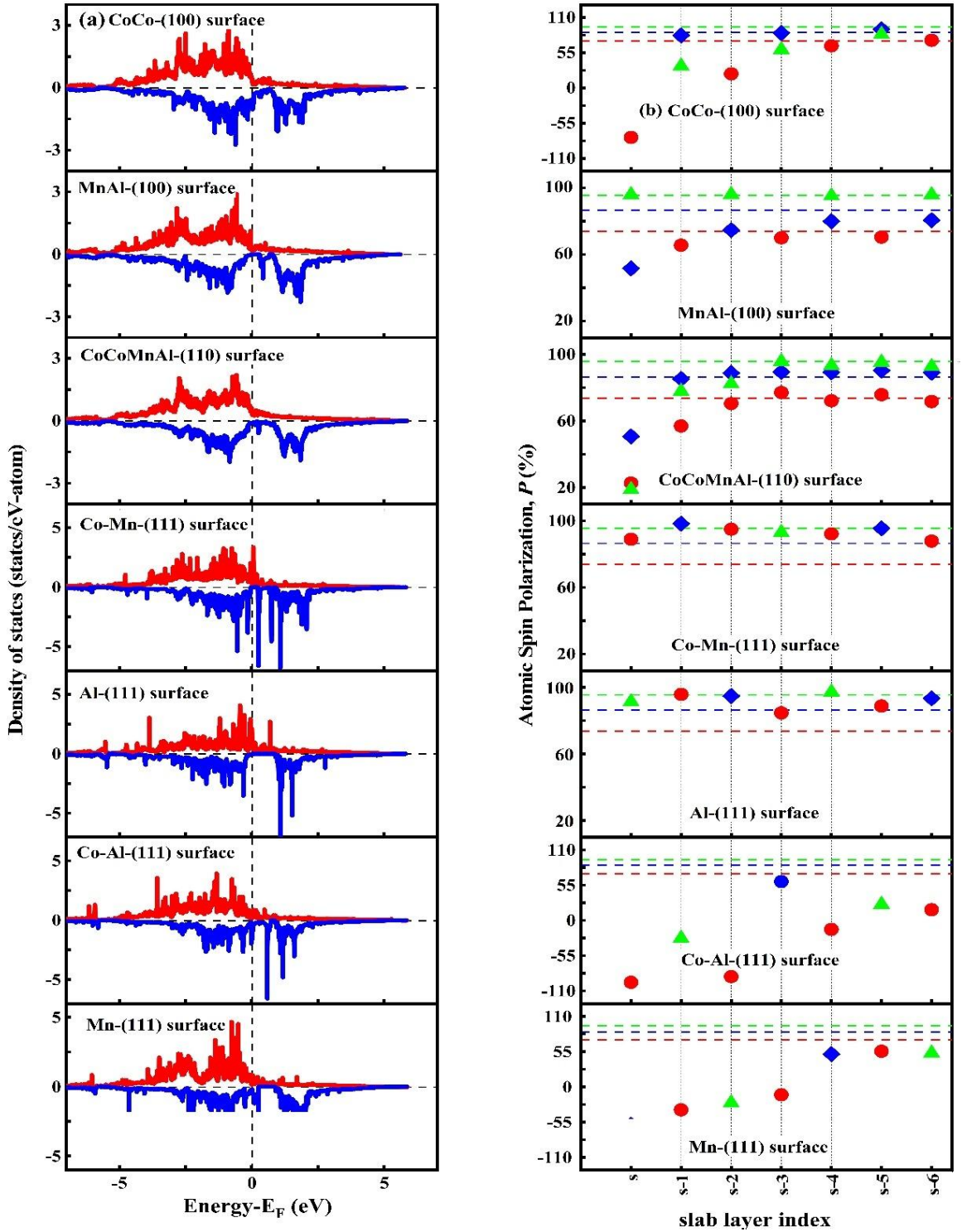

**Figure 4:** (a) Spin resolved surface density of states of various $Co_2MnAl$ surface slabs, projected exactly on top five atomic layers considering that the surface-termination effects are significant only up to the 5$^{th}$ layer. The zero energy is set to the Fermi level. (b) Atomic resolved spin polarization with respect to the slab layer index, with the corresponding spin polarization in bulk $Co_2MnAl$ indicated by dashed lines.

In order to quantify the effect of surface termination on the near-half-metallicity, the atom-resolved $P$ for all slabs at Fermi level ($E_F$) is also quantified, using the formula $P = \frac{D_1 - D_2}{D_1 + D_2} \times 100$. The values are plotted in Figure 4(b). Here, $D_1$ and $D_2$ denote the spin-up and spin-down PDOS at $E_F$. As the calculated atomic spin polarization is directly related to PDOS at $E_F$, the spin polarization for the surface slabs' atoms also varies in accordance with change in their DOS. In Figure 4(b), atomic $P$ only up to the 6$^{th}$ sub-surface (s-6) layer is given in Figure 4(b); since if further going inward the slab structures, bulk-like atomic $P$ is observed. The Al-(111) and Co-Mn-(111) surfaces show enhanced spin polarization for the near surface layers' atom, due to their improved near half-metallic electronic nature. Huge changes in atomic $P$ are observed for CoCo-(100), Co-Al-(111) and Mn-(111) surfaces, where the near surface atoms show high degree of spin polarization with the inverted signature, due to large PDOS in the corresponding minority spin channel around $E_F$. The remaining surface slabs nearly maintained the spin polarization as in bulk $Co_2MnAl$.

Also, from the analyses of atomic $P$, it is stressed here that the MnAl-(100), CoCoMnAl-(110) and Co-Mn-(111) surfaces retained the very high atomic spin polarizations for near surface atoms, like in bulk $Co_2MnAl$, despite the significant changes in their valence band and conduction band around $E_F$. Such results arise since the calculation of $P$ here accounts only for the DOS at $E_F$ and not any changes to the valence or conduction bands. Thus, the stability of half-metallicity, which is very important from the transport properties aspect, is not captured in this approach. To address this limitation, calculation of spin polarization including the electronic velocity, might be useful, as done in reference-33 [33].

As a further investigation, let us discuss the magnetization of the surface slabs. The atomic magnetic moments for surface slabs can also be understood quantitatively from atomic DOS, as the magnetic moment is given by $\mu_B$ × (difference between the total spin-up and spin-down electrons). Given that the spin-resolved DOS for the near surface atoms differs from the bulk one due to the surface formations; thereby, the change in atomic magnetic moments (AMMs) is also anticipated for the near surface layers' atoms. These changes in the AMMs share the same origin that give rise to PDOS variations. From Figure 5(a), notable changes in AMMs are observed for the all-surface slabs. For the Al-(111) surface, the surface magnetization decreases due to the reduced AMMs for the exterior atoms. For the remaining surface slabs, surface magnetization increases as the AMMs of the exterior atoms significantly increase, extending across several sub-layers. Among these surfaces, the CoCoMnAl-(110) surface displays the smallest increment, as the exterior atoms show only minor deviations in their magnetic moments compared to their bulk values. Thus, the CoCoMnAl-(110) surface can also be interpreted as exhibiting nearly the same magnetization as the bulk. Notably, these findings for magnetizations and spin polarization matched the align well with the trends reported by Sakuma *et al*. [46] and Makinistian *et al.* [20], who studied the electronic and magnetic properties only for the $Co_2MnAl$-(100) surfaces. They found that the MnAl-(100) surface retained the near half-metallic nature with increased magnetization, whereas the CoCo-(100) surface exhibited the strong metallic behavior accomplished by increased magnetization. These previous reports validate our results.

### 3.3 Surface Magnetic Anisotropy

As part of further investigations, magnetic anisotropy for the $Co_2MnAl$ surfaces is also examined in terms of magneto-crystalline anisotropy (MCA) energy. Magnetic anisotropy determines the preferred orientation of magnetization within material and plays a crucial role in the device design for various advanced applications. These include - high-density data storage, non-volatile memory devices, magnetic sensors and permanent magnets, etc. Thus, magnetic anisotropic materials are highly desirable for many spintronics applications, and studying magnetic anisotropy remains a cornerstone of spintronic research.

Typically, for the 3$d$-transition metal-based cubic Heusler compounds, like – $Co_2MnAl$, MCA is very small ($10^{-5}$-$10^{-6}$ MJ/m$^3$) due to the weak spin-orbit coupling (SOC) and small crystal field splitting. However, the reduced symmetry at the surface termination can induce a significantly large MCA, making its investigation essential for various surface terminations. Moreover, to the best of our knowledge, the magnetic anisotropy for $Co_2MnAl$ surfaces has not been studied so far. Thereby, in the present study, the MCA for $Co_2MnAl$ surfaces are studied. The calculation methods, as well as the corresponding results for MCA are as follows-

For the calculation of MCA of surface slab, which primarily arise from the broken symmetry at the slabs ends, the magnetic force theorem is adopted as implemented in Quantum Espresso package [29,47–49]. According to the magnetic force theorem, the self-consistent field calculation (SCF) is first performed for the optimized surface slabs without SOC and using a dense *k*-point grid. After that, non-self-consistent field calculations (NSCF) are performed for the magnetic moments aligned along [100] and [001] axes of slabs, using the charge density obtained from the SCF calculations along with SOC as perturbations. Here, to accurately capture the small magnetic anisotropy energies (MAE), we use a dense k-point mesh (equivalent to the 15×15×15 *k*-points in the L2$_1$-ordered structure of $Co_2MnAl$) and a stricter convergence criteria of $10^{-12}$ Ry for energy density in SCF and NSCF calculations. Finally, MCA energy and MCA constant for all the slabs are calculated as follows:

$$E_{MCA} = (E_{band}[100] - E_{band}[001]), \quad K_{MCA} = E_{MCA}/2A \tag{5}$$

Where $E_{band}[100]$ and $E_{band}[001]$ represent the total band obatained energies from NSCF step for a given surface slab. The factor of 2 in the dominator accounts for the two symmetrical surfaces at the slab ends. Depending on $E_{MCA}$ values, the surface slabs can exhibit either perpendicular magnetic anisotropy (PMA) with a positive value of $E_{MCA}$ or the in-plane magnetic anisotropy (IMA) with negative value of $E_{MCA}$. Furthermore, to identify the atomic and orbital contributions, the atomic and orbital dependent MCA is calculated according to the given equations:

$$E(MCA_\alpha) = \int^{E_F}(E - E_F)n_\alpha^{[100]}(E)dE - \int^{E_F}(E - E_F)n_\alpha^{[001]}(E)dE; \quad E(MCA_\alpha) = \sum_\beta E(MCA_{\alpha\beta}) \tag{6}$$

These contributions are plotted in Figure 5(b) and Figure S2 in the supplemental material. Here, it is important to note that for the slabs containing multiple atoms of the same kind in an atomic layer (CoCo-(100), MnAl-(100), and CoCoMnAl-(110) slabs), contribution from only one atom of each type in an atomic layer is shown in Figures. 5(b) and S2, as all the atoms of the same kind within a specific atomic layer is found to contribute similarly. For example, in the CoCo-(100) surface, there are alternating atomic layers containing either two Co atoms or a combination of Mn and Al atoms. Then, for the atomic layers containing CoCo-atoms, the contribution from only one cobalt atom is shown, as both Co atoms contribute similarly; whereas for the layers containing Mn and Al, the contributions from both Mn and Al atoms are presented, as both atoms are distinct.

As illustrated in Figure 5 (b), MCA in the surface slabs is predominantly contributed by near-surface atoms, while the inner-layers atoms show nearly magnetic isotropic nature. Specifically, for the {100}, {110}, and {111} surfaces, significant atomic-MCA is observed up to the (S-2), (S-1), and (S-4) atomic-layers, respectively. Among these atomic layers too, the primary contributions originate from the surface atoms and gradually decrease as we go inside the slab (or inward the slab direction). The remaining inner-layer atoms, beyond the (s-4) layer, display negligible MCA compared to outer layers, due to opposing orbital contributions that tend to offset the atomic contribution (as seen in Figure S2 of supplemental material). This observation that the MCA in the surface slabs primarily originates from near-surface atoms, while the inner layers show the bulk-like (magnetic isotropic) behavior - aligns well with other surface phenomena such as structural relaxations, spin polarization, and magnetic moments, where the near surface atoms play a key role. Additionally, Al-atoms contribute very little to the MCA across all slabs, irrespective to their position in slab. The layer-dependent atomic and orbital contributions to MCA, along with the total MCA for specific slabs, are as follows: -

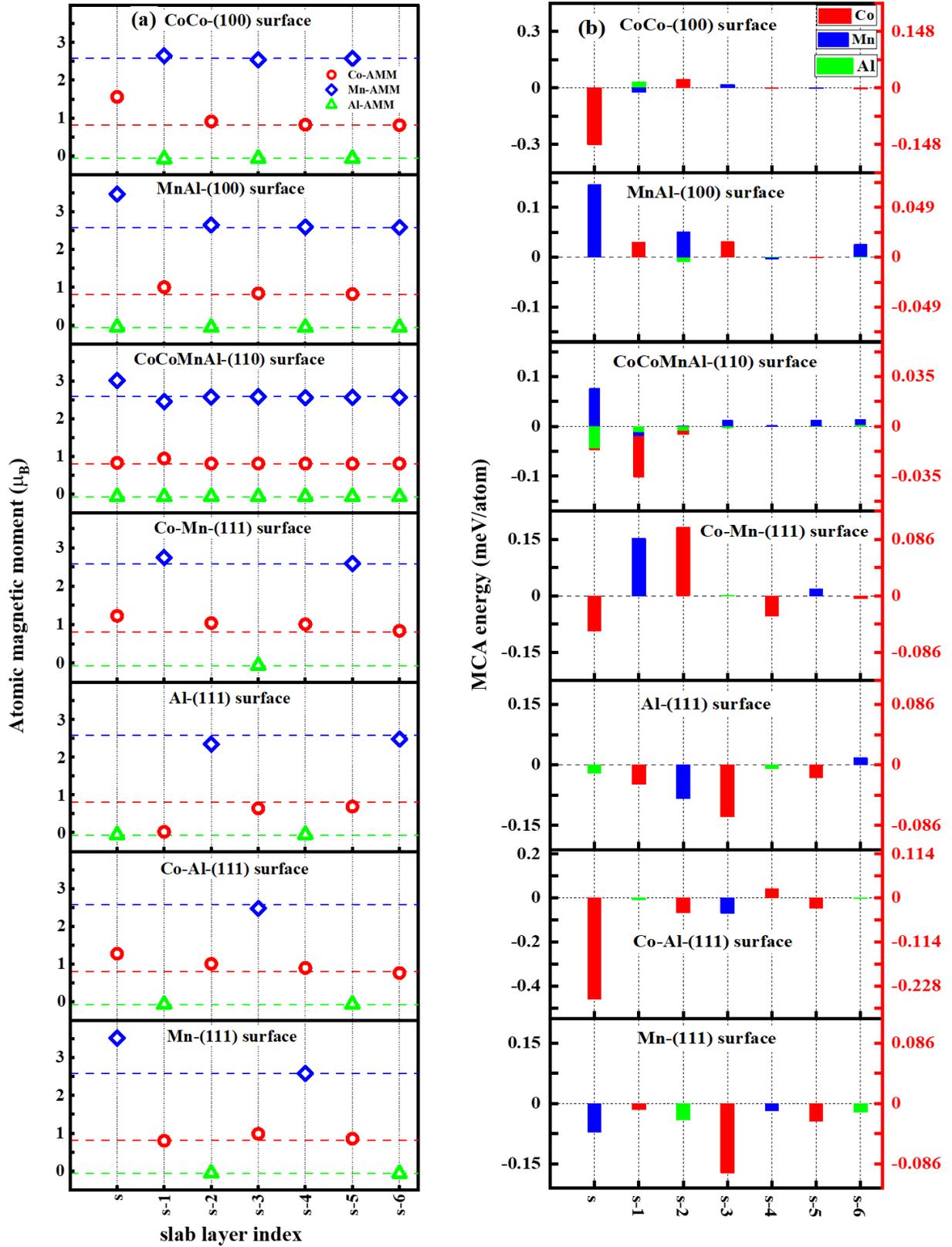

**Figure 5:** (a) Atomic magnetic moment w.r.t slab layer index, with the corresponding bulk values given by dashed lines. (b) The atomic-resolved magneto-crystalline anisotropy (MCA) energy (meV/atom) and MCA constant (mJ/m$^2$) for slab atoms with respect to slab layer index.

**(a) CoCo-(100) Surface:** The CoCo-(100) surface exhibits a negative total MCA of -0.43 mJ/m², primarily contributed by the Co surface atoms. As seen in Figure S2, the surface atomic orbitals Co-($d_{z^2}$) and Co-($d_{zy}$) contribute negatively to the MCA energy, whereas Co-($d_{xz}$), Co-($d_{x^2-y^2}$), and Co-($d_{xy}$) contribute positively. However, the dominant contributions from Co-($d_{z^2}$) and Co-($d_{zy}$) result in an overall negative MCA for the Co surface atom. The remaining atoms in slab contribute very little to MCA compared to surface atoms, leading to a net negative MCA for the CoCo-(100) slab.

**(b) MnAl-(100) Surface:** The MnAl-(100) surface exhibits a positive MCA of 0.23 mJ/m². For the Mn surface atoms, the $d_{xy}$, $d_{x^2-y^2}$, $d_{xz}$, and $d_{yz}$ orbitals contribute positively to MCA, while the $d_{z^2}$ orbitals contribute negatively. However, the net MCA remains positive for Mn surface atoms due to the overall stronger positive contributions. The first and second subsurface atoms (Co and Mn) also contribute positively to MCA, resulting in a net positive MCA for the MnAl-(100) surface.

**(c) CoCoMnAl-(110) Surface:** For the CoCoMnAl-(110) slab, Mn surface atoms exhibit positive MCA. However, negative contributions from Co and Al surface atoms, along with a significant negative contribution from the subsurface layers, result in an overall negative MCA for the (110) slab. Consequently, the CoCoMnAl-(110) slab has a total negative MCA of -0.22 mJ/m².

**(d) Co-Mn-(111) Surface:** The Co surface atoms of the Co-Mn-(111) surface exhibit negative MCA. However, positive contributions from subsurface Mn (s-1) and Co (s-2) atoms result in an overall positive MCA, reaching a maximum value of 0.33 mJ/m²—the highest among all studied slabs. The surface-Co-($d_{xy}$), surface-Co-($d_{xz}$), surface-Co-($d_{x^2-y^2}$), 1st subsurface-Mn-($d_{x^2-y^2}$), and 1st subsurface-Mn-($d_{xy}$) orbitals are the major contributors to the positive MCA (see Figure S2).

**(e) Al-(111), Co-Al-(111), and Mn-(111) Surfaces:** For the other (111) surfaces, negative atomic-MCA contributions from the near-surface atoms lead to an overall negative MCA. The calculated MCA values are -0.45 mJ/m², -0.98 mJ/m², and -1.05 mJ/m² for the Al-(111), Co-Al-(111), and Mn-(111) surfaces, respectively. The atomic- and orbital-resolved contributions to MCA can be observed in Figures 5(b) and S2.

Thus, based on these findings for MCA, it can be concluded that the MnAl-(100) and Co-Mn-(111) surfaces exhibit positive MCA (or perpendicular magnetic anisotropy), whereas the remaining surfaces exhibit in-plane magnetic anisotropy. This suggests that the growth of Co$_2$MnAl on (100)-oriented substrates (e.g., Si, GaAs, and MgO) and (111)-oriented substrates (e.g., CdS and PbS) can result in high PMA for the Co$_2$MnAl surfaces in experiments (assuming that the growth of Co$_2$MnAl on (100)-oriented and (111)-oriented substrates promotes the formation of MnAl-(100) and Co-Mn-(111) surfaces via lattice vector matching).

## 4. Conclusion

The structural, electronic, and magnetic properties of the $Co_2MnAl$ surfaces, with various atomic terminations for the (100), (110) and (111) orientations, have been studied using the plane-wave pseudopotential method within density functional theory. Among these surfaces, due to their least symmetric geometries, the {111} surfaces exhibit the most significant reconstructions for the near-surface atomic layers. On the other hand, the {100} and {110} surfaces show minor relaxations. Across all surface configurations, the MnAl-(100), Co-Al-(111) and Al-(111) surfaces show the negative surface energies, suggesting their large structural stability and spontaneous growth in experiments. Concerning half-metallicity, the MnAl-(100), CoCoMnAl-(110), Al-(111), and CoMn-(111) surfaces show the near half-metallic nature; while other surfaces- CoCo-(100), Mn-(111) and CoAl-(111) - display a strong metallic nature. Spin-polarized STM studies can help to verify these predictions. The impact of surface formation on electronic properties is further analyzed in terms of spin polarization and magnetic moments. Interestingly, for the all-surfaces with the metallic nature (CoCo-(100), Mn-(111) and CoAl-(111) surfaces), the spin polarizations of the surface and near surface atoms at Fermi level invert their polarity from the bulk values, due to presence of greater minority-spin DOS compared to majority-spin DOS, around $E_F$. On the other hand, for all surfaces with nearly half-metallic nature (i.e., MnAl-(100), CoCoMnAl-(110), Al-(111), and CoMn-(111) surfaces), a high degree of spin polarizations, nearly identical to the bulk-$Co_2MnAl$, is observed for the atoms in the surface and sub-surface layers. Regarding the surface magnetization, except for the Al-(111) surface, all other surfaces show the increased magnetization, with the smallest increment observed for the CoCoMnAl-(110) surface. In contrast, the Al-(111) surface demonstrates a significant reduction in magnetization due to the diminished moments for the exterior atoms. The magneto-crystalline anisotropy (MCA) for the surface slabs is also studied using the magnetic force theorem. The MnAl-(100) and Co-Mn-(111) surfaces show positive MCA of 0.23 $mJ/m^2$ and 0.33 $mJ/m^2$, respectively, primarily contributed by the surface and sub-surface atoms, while the other studied surfaces show a negative MCA. Thus, the electronic and magnetic properties of $Co_2MnAl$ depend on the surface termination and in turn are expected to influence the overall magnetic and transport behavior of $Co_2MnAl$ in experiments. Therefore, considering surface stability, half-metallicity, and MCA, it can be concluded from these findings that the MnAl-(100) surface is a highly promising candidate for high-performance spintronics applications. Additionally, if the structural instability of Co-Mn-(111) surfaces is ignored (e.g., by growing it through non-equilibrium growth techniques), this surface morphology is also favorable for spintronics applications due to its near-half-metallic nature and large positive MCA. However, for specific spintronics applications, in addition to the surface stability, the half-metallicity and the MCA, other device-specific properties must also be considered. We anticipate further experimental advancements to validate these predictions.

Finally, one more important conclusion can be drawn from the present study. A review of the existing literature reveals consistent trends between the present study findings and other previous studies on various Co-based Heusler alloys. These studies include:

- Surface energies of Co$_2$VGa-(100) surfaces [50], Co$_2$MnSi-(100) surfaces [45], and Co$_2$VGa-(111) surfaces [38].

- Half-metallicity and magnetism of Co$_2$VGa-(100) surfaces [50], Co$_2$CrAl-(100) surfaces [6], Co$_2$MnSi [5], (100) surfaces of Co$_2$CrAl, Co$_2$MnGe, and Co$_2$MnSi Heusler alloys [51], Co$_2$CrAl-(110) surfaces [40], and Co$_2$VGa-(111) surfaces [38].

- MCA of Co$_2$CrAl-(100) surfaces [6], and Co$_2$MnSi-(100) surfaces [5].

Importantly, we found no reports of Co-based Heusler alloys exhibiting opposing trends in surface energies, half-metallicity, or magnetism. Therefore, the results in present can also contribute to establishing a generalized behaviour for studying the surface stability, half-metallicity, magnetization, and magnetic anisotropy in Co$_2$YZ Heusler alloy surfaces.

## Acknowledgements

PARAM Rudra, a national supercomputing facility, at Inter-University Accelerator Centre (IUAC), New Delhi, has been used to obtain the results presented in this paper. A.K. acknowledges Council of Scientific and Industrial Research (Grant No. 09/086(1356)/2019-EMR-I) India, for the senior research fellowship. The authors thank Sanjay Kumar Kedia (IUAC, New Delhi) for the helpful discussions.## References

# Supplemental Material

**Table S1:** The changes in the atomic position owed by atomic relaxation (absolute values and in terms of percent change relative to the interlayer distances). See text for abbreviations.

| Atomic termination-orientation | Slab layer (atoms) | Change in atomic position after relaxation, Δz (Å) | % (Δz), relative to the interlayer distance |
|---|---|---|---|
| CoCo-(100) surface | s (Co, Co) | 0.002, 0.002 | 0.140, 0.140 |
| | s-1 (Mn, Al) | -0.025, 0.010 | -1.760, 0.710 |
| | s-2 (Co, Co) | 0.015, 0.015 | 1.027, 1.027 |
| | s-3 (Mn, Al) | -0.003, 0.005 | -0.219, 0.316 |
| MnAl-(100) surfaces | s (Mn, Al) | -0.026, 0.075 | -1.814, 5.250 |
| | s-1 (Co, Co) | -0.039, -0.039 | -2.727, -2.7270 |
| | s-2 (Mn, Al) | 0.011, -0.003 | 0.780, -0.243 |
| | s-3 (Co, Co) | -0.005, -0.005 | -0.335, -0.335 |
| CoCoMnAl-(110) surface | s (Co, Co, Mn, Al) | 0.062, 0.062, -0.002, -0.060 | 3.098, 3.098, -0.086, -2.992 |
| | s-1 (Co, Co, Mn, Al) | -0.005, -0.005, -0.028, 0.015 | -0.270, -0.270, -1.401, 0.751 |
| | s-2 (Co, Co, Mn, Al) | -0.001, -0.001, -0.004, -0.009 | -0.065, -0.065, -0.214, -0.472 |
| | s-3 (Co, Co, Mn, Al) | 0.000, 0.000, -0.003, -0.001 | -0.019, -0.019, -0.146, -0.070 |
| Co-Mn-(111) surface | s (Co) | 0.095 | 9.675 |
| | s-1 (Mn) | 0.036 | 3.680 |
| | s-2 (Co) | -0.097 | -9.810 |
| | s-3 (Al) | 0.008 | 0.799 |
| Al-(111) surface | s (Al) | -0.130 | -13.154 |
| | s-1 (Co) | -0.184 | -18.654 |
| | s-2 (Mn) | 0.141 | 14.354 |
| | s-3 (Co) | -0.051 | -5.161 |
| Co-Al-(111) surface | s (Co) | -0.174 | -17.660 |
| | s-1 (Al) | -0.041 | -4.160 |
| | s-2 (Co) | 0.063 | 6.415 |
| | s-3 (Mn) | -0.007 | -0.743 |
| Mn-(111) surface | s (Mn) | 0.012 | 1.227 |
| | s-1 (Co) | -0.071 | -7.199 |
| | s-2 (Al) | 0.035 | 3.520 |
| | s-3 (Co) | 0.012 | 1.204 |

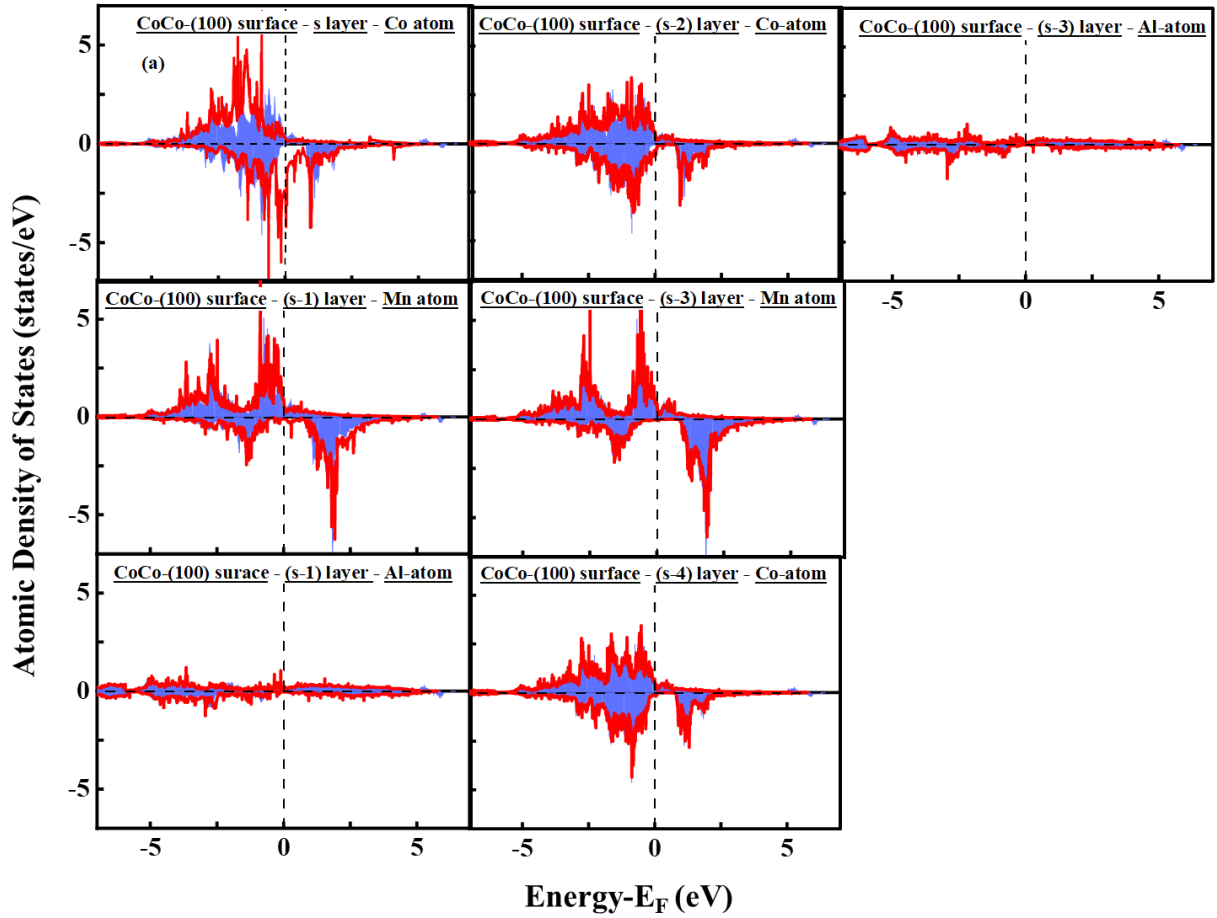

**Figure S1(a):** Atomic density of states of the near surface atoms for CoCo-(100) surface. The layer indices (e.g., s, (s-1), etc.) and atomic symbols (Co, Mn and Al) are given in the inset of each picture. The blue shaded region corresponds to the PDOS for the corresponding atom in bulk-$Co_2MnAl$.

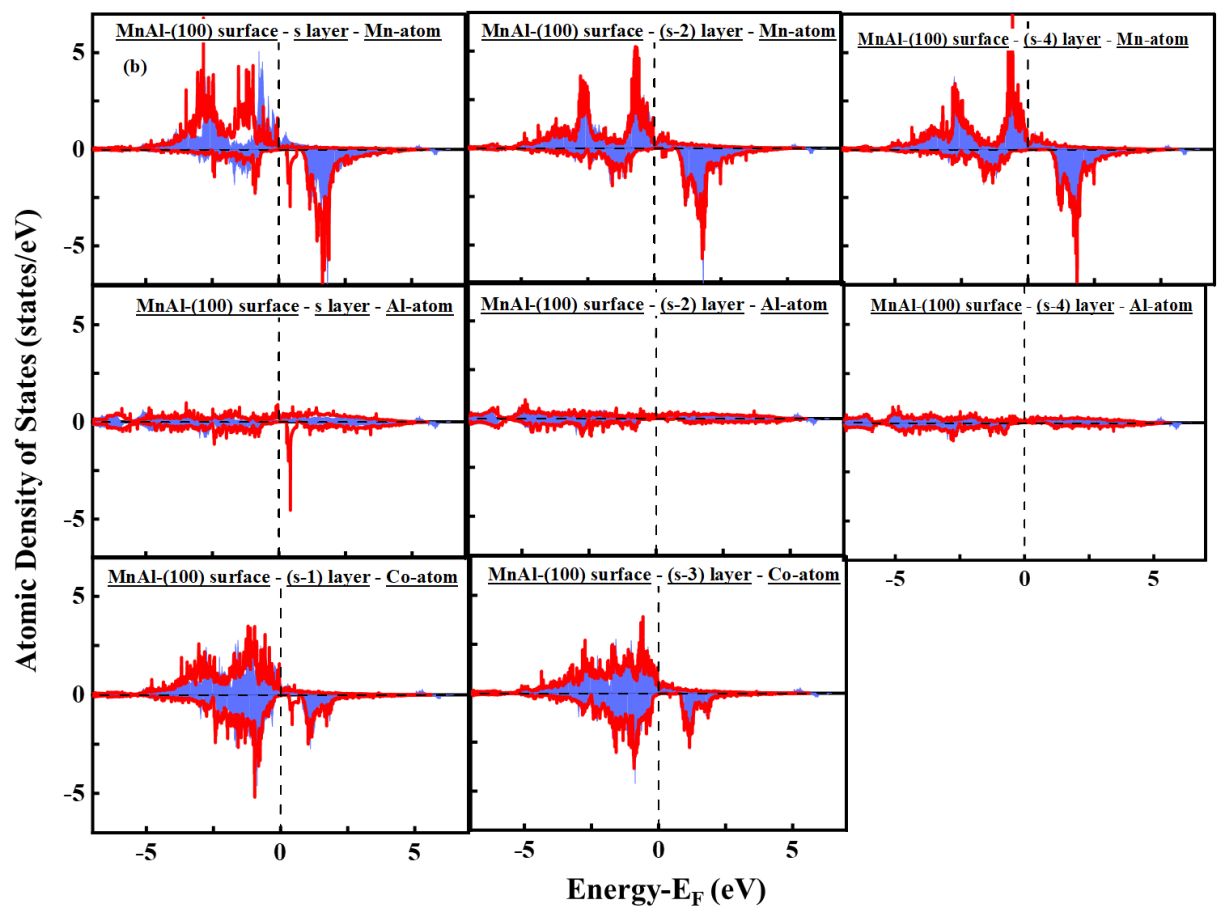

**Figure S1(b):** Same as of Figure S1(a), but for MnAl-(100) surface.

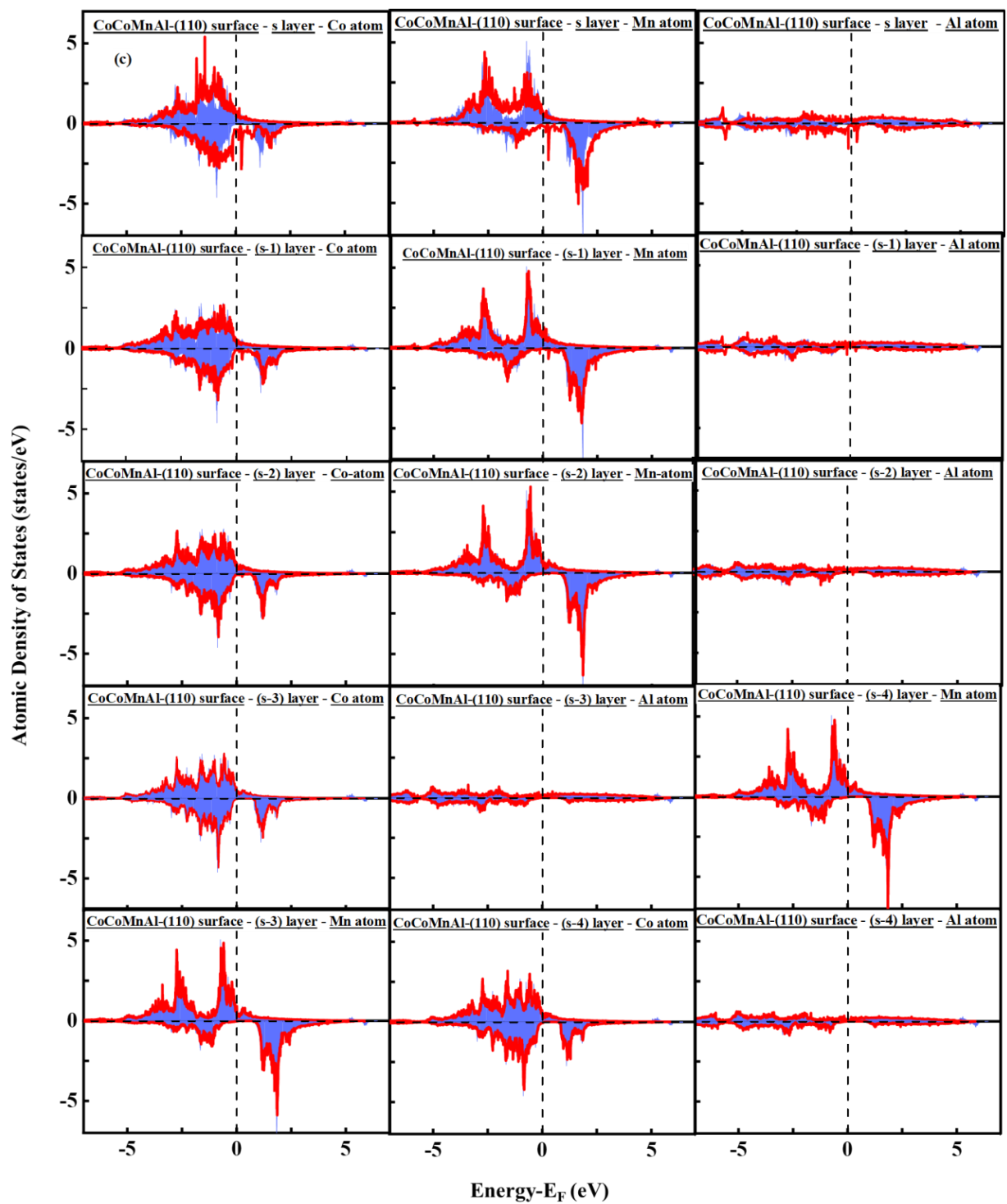

**Figure S1(c):** Same as of Figure S1(a), but for CoCoMnAl-(110) surface.

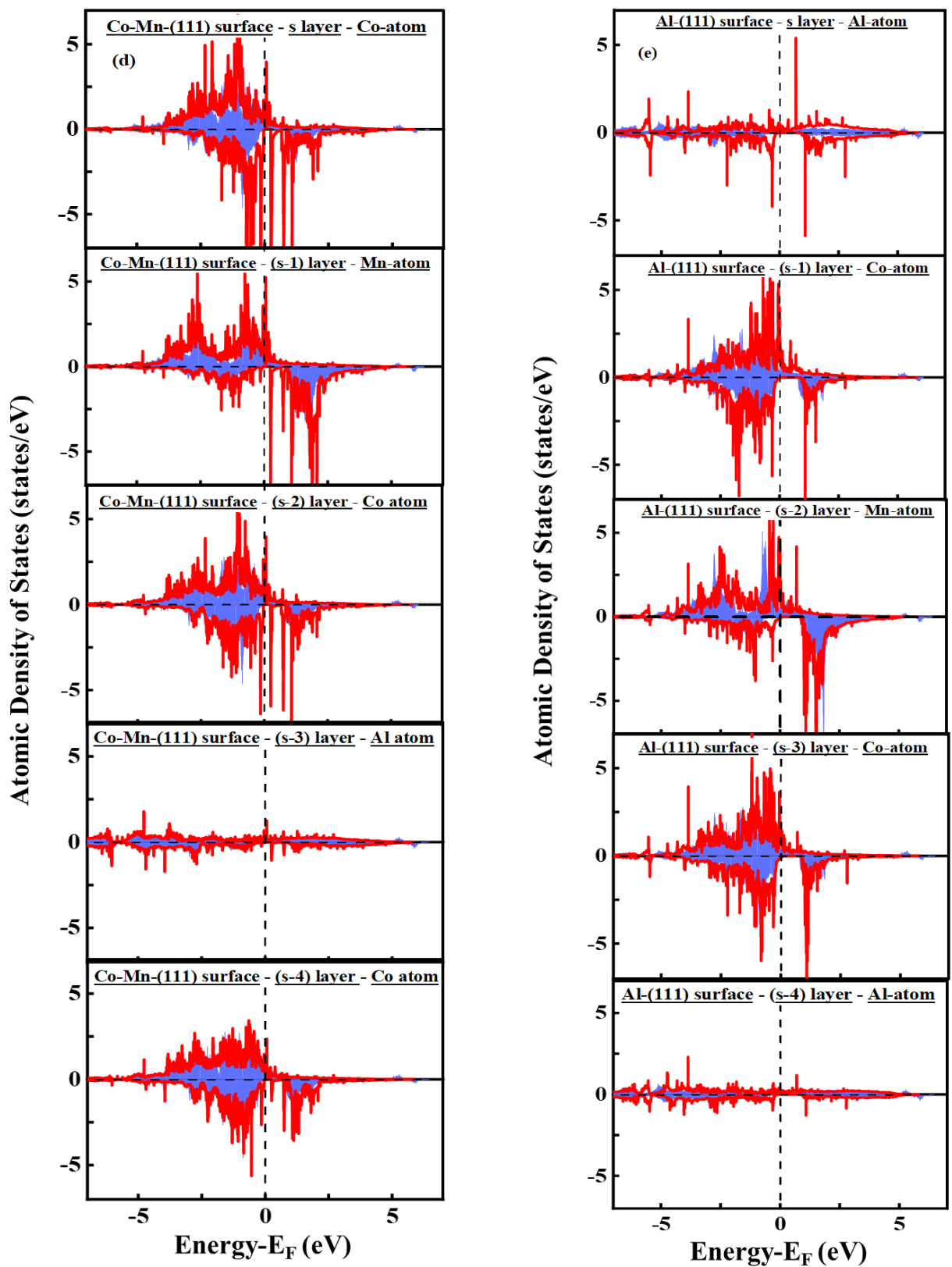

**Figure S1(d)-(e):** Same as of Figure S1(a), but for Co-Mn-(111), Al-(111), Co-Al-(111) and Mn-(111) surfaces.

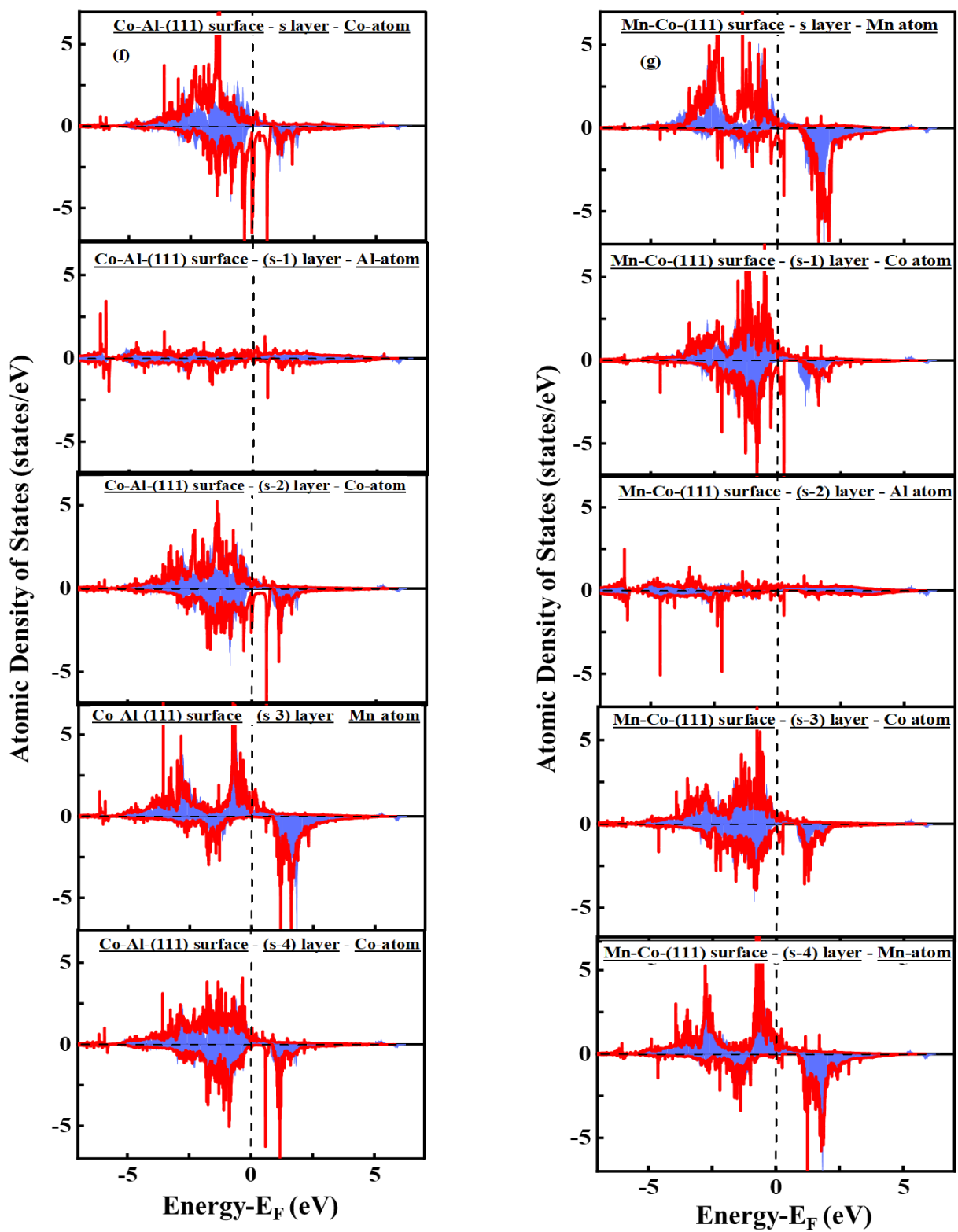

**Figure S1(f)-(g):** Same as of Figure S1(a), but for Co-Mn-(111), Al-(111), Co-Al-(111) and Mn-(111) surfaces.

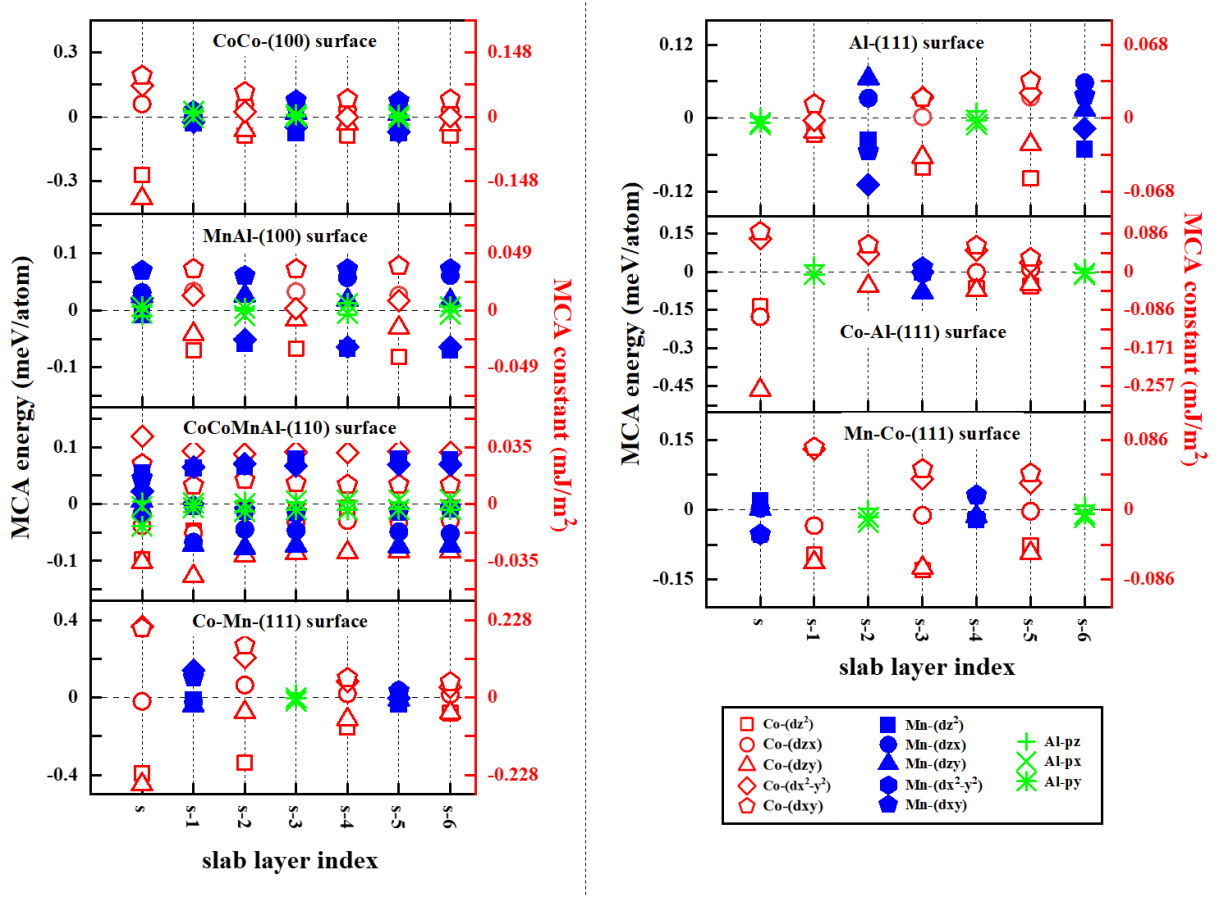

**Figure S2**: The orbital-resolved MCA for the different surface slabs with respect to the slab layer index.